\input epsf.tex  

\documentclass[preprint2]{aastex}

\newcommand{\simless}{\mathbin{\lower 3pt\hbox
     {$\rlap{\raise 5pt\hbox{$\char'074$}}\mathchar"7218$}}} 
\newcommand{\simgreat}{\mathbin{\lower 3pt\hbox
     {$\rlap{\raise 5pt\hbox{$\char'076$}}\mathchar"7218$}}} 
\shorttitle{  Imaging of Southern Young Binaries }
\shortauthors{ Koresko }

\begin{document}


\title{ Imaging the Circumstellar Environments of Young Binaries in Southern
Star-Forming Regions}


\author{ Chris D. Koresko }
\affil{ Interferometry Science Center, Caltech, Pasadena, CA 91125 }
    
\begin{abstract}

A sample of 14 pre-main sequence visual binary systems in southern
star-forming regions were imaged at 2 $\mu$m with the Keck I telescope to
search for tertiary companions, scattered-light disks, and compact nebulae
at linear scales of 5--100 AU.   Speckle holography was used to produce
images with diffraction-limited resolution and high dynamic range, and
photometry in four filters between J (1.25 $\mu$m) and L' (3.8 $\mu$m) was
used to provide a diagnostic of the infrared excess.  Of the 14 visual
binary systems studied, 9 contain components which show evidence for
resolved structure.  Two of them (WSB~18 and B59-1) have well-resolved
tertiary companions separated from the primary stars by $\sim 100$~mas. 
The remainder are only marginally resolved at the diffraction limit of the
telescope.  Sz~116, AS~205, Elias~2-22, ESO~H$\alpha$ 282, and B59-2 each
have one component which is marginally-resolved along one axis and
unresolved on the perpendicular axis, consistent with tertiaries at
separations between $\sim 6$ and 40~mas.  If all of these systems are
hierarchical triples, then fully half of the nominal binaries in the
sample are triple.  The primaries in Elias~2-49 and WSB~71 are marginally
resolved along all position angles, suggesting the presence of dust
halos.  No disks were unambiguously detected in the sample. 

\end{abstract}


\section{ Introduction }

It has recently become clear that the great majority of the young low-mass
stars in the nearest active star-forming regions (SFRs) are found in binary
or multiple systems (Leinert et al. 1993; Ghez et al. 1993; Ghez et al.
1997).  Although there is evidence that the mass in circumstellar material
is smaller in binaries with separations $\simless 100$ AU than in wider
binaries or single stars (Jensen, Mathieu, \& Fuller 1996), a majority of
young binaries emit excess radiation at infrared or submillimeter
wavelengths which is typically presumed to arise in dusty disks.  If the
presence of one or more stellar companions does not severely hamper a disk's
ability to produce a planetary system, then binary and multiple systems
probably contain the majority of the planets in the nearby SFRs. 

At present, only two multiple-star systems are known to contain
well-ordered, edge-on circumstellar disks visible because of their occlusion
of the central star.  The HK~Tauri~B and HV~Tauri~C disks exhibit a
morphology much like that predicted by scattered-light models: A pair of
bright lenslike reflection nebulae trace the upper and lower surfaces of the
disk, while the dense region nearer the midplane is dark (Stapelfeldt et al.
1998; Koresko 1998; Monin \& Bouvier 2000).  There is no central point
source seen in either the visible or near-infrared images.  The
near-infrared color of the HK~Tauri~B star+disk system is bluer than that
of the primary star, presumably because the light we see is scattered by
small particles.  

A handful of visual binary companions to T~Tauri stars are ``infrared
companions" (IRCs), objects unusually faint at visible wavelengths but
bright in the infrared.  The IRCs are mysterious because their spectral
energy distributions show strong reddening and infrared excess reminiscent
of stars at younger evolutionary phases than their primaries ({\it e.g.,}
Koresko, Herbst, \& Leinert 1997).  The structures seen in the two IRCs
which have been resolved in published high-resolution images are not edge-on
disk systems, but a diffuse and irregular nebula (Koresko et al. 1999) and a
close double (Koresko 2000).  Among the present sample, the companion to
Elias~2-22 is known from previous observations (Chelli et al. 1988) to be an
IRC.

Clearly, the circumstellar environments of young binary stars exhibit a
range of phenomena which are accessible to modern imaging techniques.  This
paper describes the first results of an ongoing effort to characterize the
circumstellar environments in a larger sample of young binary systems.  The
sources were chosen from the binaries detected in the visible-light imaging
survey of Reipurth \& Zinnecker (1993; hereafter RZ).  They include all the
binaries, with the exception of Sz~120, with separations between 1\arcsec.0
and 3\arcsec.7 and right ascensions between $16^{\rm h}~06^{\rm m}$ and
$17^{\rm h}~10^{\rm m}$ (equinox 1950). This RA range includes all the RZ
binaries in the Ophiuchus SFR, and several additional groups, for a total of
14 systems.  The range of separations was chosen to avoid having the
seeing-limited Point-Spread Functions (PSFs) overlap or fall off the edges
of the detector.  With one exception, the distances to the targets are given
by RZ as 160~pc.  Their physical separations are larger than 100~AU, so they
are not expected to be strongly depleted of disk material via disk-companion
interactions (Jensen, Mathieu, \& Fuller 1996).  

The sources are a somewhat mixed bag, consisting mainly of T~Tauri stars
but including at least one Herbig~Ae/Be star (Elias~2-49) and a likely
Post-T~Tauri star (WSB~4).  Many of the sources are well-studied, but a
handful lack even published spectral types.  Ten of the 14 systems are
members of RZ's ``systematic survey sample", which for the present purpose
means that they were known to be members of dark clouds $\sim 160$ pc
distant and are included in a list of young stars selected on the basis of
a youth indicator such as near- or far-infrared excess or Ca II HK line
emission.  The remaining four systems are AS~205, both of whose components
are Classical T~Tauri stars (CTTS) (Prato et al. 1997); Elias~2-26,
another CTTS (Martin et al. 1998); Elias~2-49, a Herbig Ae/Be star in
Hillenbrand's Group~I (Hillenbrand 1992); and ESO H$\alpha$~282, about
which little seems to be known beyond the fact that it is an emission-line
source.

The 14 pre-main sequence binary systems were observed using speckle
holography at near-infrared wavelengths $\sim 2~\mu$m to produce
diffraction-limited images with dynamic range of $\sim 2000$ at 0\arcsec.3
from the star, which is much better than would have been possible with
standard speckle imaging.  The improved dynamic range is possible because
the PSF is measured simultaneously with the science observations. 
Near-infrared photometry in the J, H, K, and L' bands was obtained the
following night to measure the infrared excess of each component as a tracer
of warm dust. 

Two new tertiary companions with separations $\sim 0$\arcsec.1 were
detected, and an additional 7 objects were found to be marginally resolved. 
Of these, 5 are extended along only one direction, which is consistent with
the presence of a very close tertiary companion or a narrow extended
structure such as a compact edge-on disk.  The remaining two
marginally-resolved objects were extended in all directions, which would be
consistent with extended nebulosity but not with tertiary companions.  

No new edge-on disks were unambiguously detected.  It is known from
theoretical models ({\it e.g.,} Wood et al. 1998) and from recent
high-resolution images ({\it e.g.,} Burrows et al. 1996) that pre-main
sequence circumstellar disks with fiducial mass distributions are readily
detectable in visible-light or near-infrared images only when they are
viewed along a line of sight close to the disk midplane.  In this
orientation, the direct radiation from the central star is blocked, making
it possible to see the much fainter scattered light which traces the disk
surface. The range of viewing angles over which the disk hides the star, and
therefore the likelihood of detecting it if its orientation is random,
depends on the disk's mass, radius, and flaring, and potentially on the
degree to which the scattering dust has settled toward the midplane.

%
%
%

\section{ Observations }

\subsection{ Holographic Imaging }

The holography data were taken at the 10~m Keck~1 telescope on 1999~June~1
using the Near-Infrared Camera (NIRC; Matthews \& Soifer 1994) with the
NIRC Image Converter (Matthews et al. 1996) to produce a magnified pixel
spacing of 20 mas, approximately Nyquist-sampling the $\sim 50$ mas
diffraction limit at the $\sim 2~\mu$m observing wavelength.  The field of
the $256\times 256$ pixel detector was 5\arcsec.1.  The optical bandpass
was set using either a standard K photometric filter, or, for stars which
would saturate the detector in the K band, one of the narrower Ks or CH4
filters.  The observations of each visual binary consisted of hundreds of
frames with an integration time of  0.14 sec per frame.  This short
exposure time partially ``froze" the atmospheric seeing, so that the raw
PSF consisted of distinguishable speckles.  The separations of these
visual binaries were large enough compared to the 0\arcsec.5 typical
seeing that only a small fraction of the frames were rejected due to poor
separation of the stars.  The data for each visual binary system were
divided into $\sim 7$ series of frames, with $\sim 140$ frames in each
series.  Series taken on blank sky were used to estimate the background
and bias level.  A journal of the holographic observations is given in
Table \ref{HoloObsJournal}.

Individual frames were calibrated by subtracting mean sky frames, dividing
by flatfield images, and ``fixing" bad pixels.  A model was computed for the
detector ``bleed" signal which extended along the readout direction, as
described in the NIRC manual, and this was subtracted from the calibrated
frame.  For each frame, a subframe centered on each of the stars was
extracted, with one subframe being treated as the science measurement and
the other serving as a measurement of the instantaneous point-spread
function (PSF).  The subframes were either $64\times 64$ or $48\times 48$
pixels in size; the smaller subframes were necessary for doubles whose
separations were less than 64 pixels.  The Fourier power spectra of the
target and PSF subframes, and the cross-spectrum ({\it i.e.,} the Fourier
transform of the cross-correlation) of the two subframes, were computed for
each input frame.  These quantities are meaningful within a roughly
hexagonal region of the $(u,v)$ plane containing spatial frequencies below
the diffraction limit of the hexagonal Keck primary mirror.    

\begin{deluxetable}{lllllll}
\footnotesize
\tablecaption{ Holographic Observations 
\label{HoloObsJournal}}
\tablewidth{0pt}
\tablehead{
\colhead{Object} & \colhead{Alt. Name} & \colhead{Region}  & \colhead{D(pc)} & \colhead{Filter} &
\colhead{Frames} & \colhead{Subframe Size}
} 
\startdata
Sz 116						& HBC 625		 &	Lupus III	& 150	 &   K   &  887     &  64             \\
AS 205						& HBC 254		 &	B 40			& ?		 &   CH4 &  941     &  64             \\
WSB 4 						& 					 &	Oph				& 160	 &   K   &  1014    &  64             \\
WSB 18						& 					 &	Oph				& 160	 &   K   &  868     &  48             \\
WSB 19            & 					 &	Oph				& 160	 &   K   &  870     &  64             \\
WSB 26						& DoAr 23		 &	Oph				& 160	 &   K   &  848     &  48             \\
Elias 2-22			  & DoAr 24E	 &  Oph 			& 160	 &   Ks  &  716     &  64             \\
WSB 35            & 					 &	Oph				& 160	 &   K   &  864     &  64             \\
Elias 2-26				& ROX 15		 &	Oph				& 160	 &   K   &  857     &  64             \\
WSB 71						& 					 &	Oph 			& 160	 &   K   &  813     &  64             \\
Elias 2-49				& HD 150193	 &	Oph				& 160	 &   Ks  &  384     &  48             \\
ESO H$\alpha$ 282 & 					 &	L 162			& 160	 &   K   &  848     &  48             \\
B59-1 						& 					 &	B 59			& 160	 &   K   &  822     &  64             \\
B59-2 						& 					 &	B 59			& 160	 &   K   &  858     &  64             \\
\enddata
\tablecomments{ 
Holographic observations of the 14 pre-main sequence visual binary
systems.  The filters' center wavelengths and half-power bandwidths,
measured in microns, are as follows: K ($2.21 \pm 0.42$), Ks ($2.16 \pm
0.33$), CH4 ($2.27 \pm 0.16$).  Distances are from Reipurth \& Zinnecker 1993.
}
\end{deluxetable}

The cross-spectrum, the target power, and the PSF power were averaged over
the whole series of good frames.  An averaged Fourier power spectrum and
Fourier phase for the target, corrected for the distortions due to the
atmosphere and the telescope, were reconstructed from them as follows:  The
power spectrum was estimated as the ratio of the average target power to the
average PSF power, after removing a noise bias term from each.  The phase
was estimated as the phase of the averaged cross-spectrum.  Uncertainties in
the power spectrum and phase were estimated from the scatter between the
averages computed for the $\sim 7$ series for each visual binary system. The
PSF star in each binary was chosen arbitrarily, and the choice was reversed
if it showed signs of resolved structure.  In no system was there clear
evidence that both stars were resolved, although in principle the technique
would be insensitive to this if the morphologies of the two stars were
sufficiently similar.

The reconstructed Fourier transforms were compared with models of single
stars, double stars, and elliptical Gaussian halos.  These choices clearly
cannot span the range of all possible {\it astrophysical} models, but for
targets which are only marginally resolved, they are sufficient to fit
the features in the data. There is a degree of degeneracy in these model
fits, in the sense that an unresolved star can always be fitted by a
double-star model (with either zero separation or zero brightness ratio) or
by a halo model (with zero axis lengths).  Further, an object which is
marginally resolved along only one axis can be fitted by either a
double-star model or a halo model with a zero semiminor axis, since either
model produces a power spectrum having a single broad stripe, and a nearly
featureless phase, in the measured frequency range.  

The comparisons with models were done separately for each series of
frames.   Only frequencies below that for which the median uncertainty rose
to four times its low-frequency value at were used.  For the double-star and
halo models, the fitting was done 11 times for each series, with the
starting values of the parameters varied randomly, and the fit with the
smallest $\chi^2$ value was chosen to represent that series.  This was
necessary because for some starting parameters the fitting routine became
trapped in local minima in the $\chi^2$ surface.  The final parameter values
for each model were the averages of the results for the series, and the
quoted uncertainties are the standard deviations of the means of those
results.  These statistical uncertainties are likely to be optimistic
because they do not include the effects of systematic biases.  The
estimation of physical parameters is rather delicate and easily affected by
those biases, especially for the marginally-resolved sources. 

For the marginally-resolved sources, the final misfit ($\chi^2$ per degree
of freedom) for the power spectrum was close to unity.  The power spectra
are more useful than the phases for marginally-resolved systems because the
power spectra begin to show strong parameter-dependent structure at lower
spatial frequencies than the phases do.  On the other hand, the position
angles derived from power spectra suffer from an ambiguity of 180 deg ({\it
e.g.,} it is not possible to tell which star in a double is brighter).  For
the fully-resolved sources, which in the present sample are both
primary-tertiary doubles, the fits to the power spectra tend to be poorer. 
This may be because the power spectra, being positive-definite, are
susceptible to biases caused by zero-mean noise, while the phase tends to be
unbiased by it.  The phase fits also produce unambiguous position angles for
resolved doubles.  In summary, the phase data are more useful for resolved
doubles, while the power spectra are more useful for marginally-resolved
objects.




\subsection{ Near-Infrared Photometry }

The photometric fluxes of the target visual binary systems were crudely
measured in the J, H, K, and L' filters at the Keck~I telescope on
1999~June~2, by imaging them with NIRC.  For the J, H, and K filters,
subframes of $256\times128$ pixels were used with the Image Converter, while
for the L' filter the Image Converter was removed and the subframe size was
reduced to $32\times 32$ or $64\times 32$ pixels.  These choices were driven
largely by the need to read out the detector fast enough to avoid saturating
it on the brighter stars or, at the longest wavelength, on the blank sky. 
The stars determined the peak signal levels in the J, H, and K filters, so
the exposure times were adjusted between 68~msec and 4~sec, and no coadding
was done.  For the L' filter the frames consisted of 1000 coadds of 5~msec
each.  A total of 10 frames were taken on each target in each filter. 
Periodically, frames were obtained on blank sky to estimate the background
level.  SR3 served as a photometric standard for the J, H, and K filters,
and HD~161793 was used for the L' filter.

The frames were calibrated in the usual way, with sky-subtraction,
flatfielding, and pixel-fixing.  To derive the fluxes in the frames, each
binary was fitted using a model in which the two stars were represented by
circular Gaussians of adjustable size, intensity, and position.  Although
the Gaussians were not very good models for the PSFs, the fact that the PSF
in each image was essentially identical for the two stars permitted the
ratio of their brightnesses and the distance between them to be estimated
accurately using this technique.  The brightness ratios were combined with
total flux measurements to derive the fluxes for the individual stars.  At
J, H, and K, these total fluxes were computed by summing the values of all
pixels within 32 pixels of either star, after subtracting the background
level estimated as the median value of all pixels at least 64 pixels away
from either star.  In the L' filter the total flux was simply the sum of all
pixel-values, after subtraction of a background level.  The photometric
fluxes were calibrated in the usual way by comparison of these numbers with
similar results on standard stars.  No attempt was made to correct for the
differences in airmass, as those corrections were found to be smaller than
the statistical uncertainties.  The final uncertainties in the magnitudes
are approximately 0.2 mag.  

Separations and position angles of the binaries were derived from the model
fits in each filter.  The platescale with IC out of the beam has been
carefully measured by the Keck Observatory staff, and the pixel spacing is
quoted in the instrument manual as 0\arcsec.1500.  By contrast, the
0\arcsec.02 nominal platescale with the IC in is relatively uncertain.  On
the other hand, the measurements of the separation vectors taken with the
smaller pixels are likely to be significantly more precise, in the sense of
having smaller uncertainties if the platescale error is neglected.  To take
advantage of this, the assumed platescale in each of the J, H, and K bands
was adjusted to make the mean separation for all the visual binaries (except
for Sz~116, which was not measured in the L' filter, and Elias~2-49, which
is anomalous as discussed below)  equal to the mean separation in the IC-out
L' images.  The required IC-in pixel spacing of 0\arcsec.0207 was consistent
between the J, H, and K images.  Similarly, the assumed orientation of the
IC-in images was adjusted to make the mean difference between the IC-in and
IC-out position angle measurements vanish.

\section{ Results and Discussion }

\begin{figure*}[htbp]
\epsfxsize=16.5cm
\epsfbox{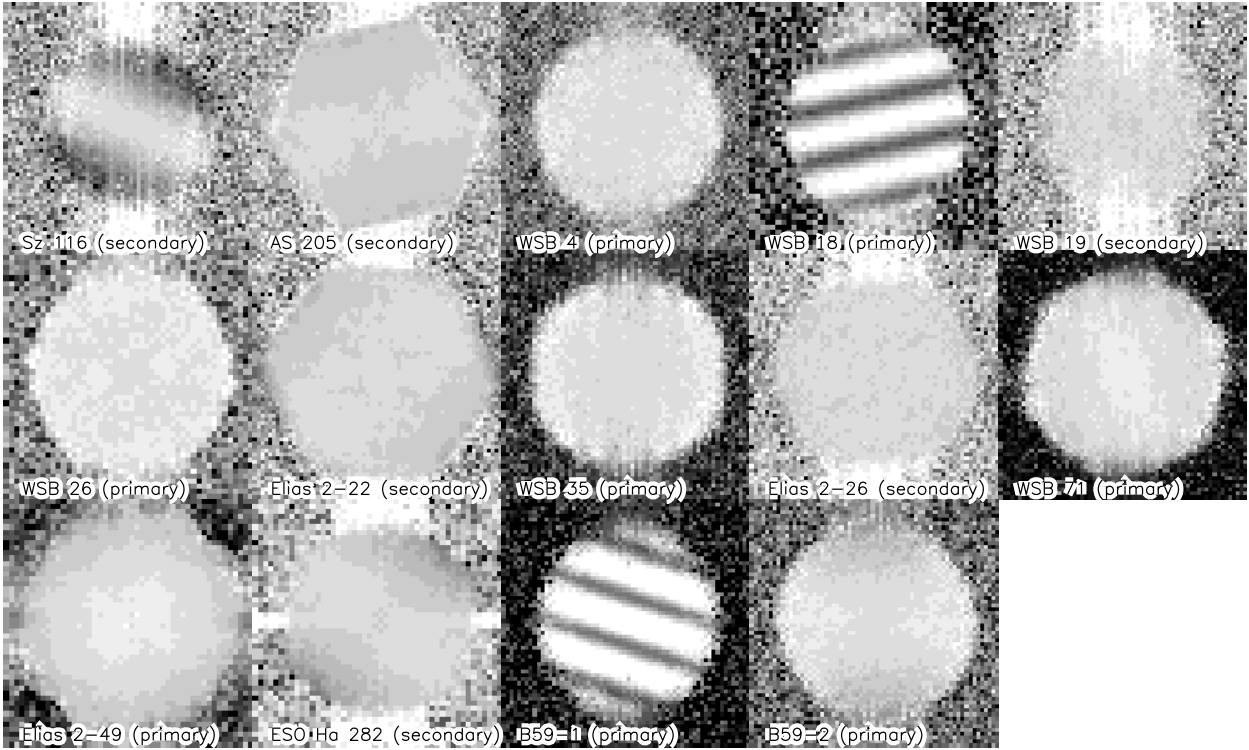}
\caption[ ]{ Reconstructed visibility amplitudes for the target components
in the 14 visual binaries, displayed with the zero frequency at the center
of each plot, and in the orientation natural to the detector. The measured
amplitudes extend from the zero-frequency point to something close to the
diffraction limit of the Keck primary mirror, whose hexagonal shape is
visible in the data on the brighter stars.  At high frequencies detector
noise produces stripes parallel to the horizontal or vertical axes. The
striped structure consistent with the presence of close tertiary
companions is obvious in Sz~116, WSB~18, and B59-1.  Closer inspection
reveals similar (but broader) central stripes in AS~205,
ESO~H$\alpha$~282, B59-2, and possibly in Elias~2-22.  The amplitudes of
WSB~71 and Elias~2-49 decrease in all directions, which is inconsistent
with a tertiary-star model. }
\label{Power2D}
\end{figure*}

The results of the holographic imaging of the target binaries are summarized
in Table \ref{HoloSummary}, and the Fourier power spectra are displayed in
Figure \ref{Power2D}.  The Table presents the values of $\chi^2$ per degree
of freedom ({\it i.e.,} per frequency element in the $(u,v)$ plane) for 
unresolved-star, double-star, and halo models of the target in each of the
14 visual binary systems studied.  For the two fully-resolved doubles, the
fits were made to the Fourier phase, resulting in unambiguous position angle
measurements, while for all other sources the fits were made to the power
spectra.  In the following discussion, a target is considered to be resolved
if either the double-star or the halo model improves the $\chi^2$ value by
more than 0.5 compared to the unresolved-star model.  This criterion
provides a clean separation between the unresolved and extended targets.

\bigskip

\begin{deluxetable}{llclllc}
\footnotesize
\tablecaption{ Summary of Holographic Imaging Results 
\label{HoloSummary}}
\tablewidth{0pt}
\tablehead{
\colhead{Object} & \colhead{Fit To} & \colhead{Target} & \colhead{$\chi^2_{u}$} & \colhead{$\chi^2_{d}$} & \colhead{$\chi^2_{h}$} & \colhead{Comment}
} 
\startdata
Sz 116            & power & sec & 9.4 & 1.5  & 1.5  &  1   \\  
AS 205            & power & sec & 4.7 & 1.1  & 1.2  &  1   \\  
WSB 4             & power & pri & 1.1 & 1.1  & 1.1  &  4   \\  
WSB 18            & phase & pri & 96  & 1.2  & 96   &  2   \\  
WSB 19            & power & sec & 1.6 & 1.6  & 1.6  &  4   \\  
WSB 26            & power & pri & 1.1 & 1.1  & 1.1  &  4   \\  
Elias 2-22        & power & sec & 1.9 & 1.1  & 1.1  &  1,5 \\  
WSB 35            & power & pri & 1.4 & 1.4  & 1.4  &  4   \\  
Elias 2-26        & power & sec & 1.1 & 1.1  & 1.1  &  4   \\  
WSB 71            & power & pri & 2.1 & 1.5  & 1.3  &  3   \\  
Elias 2-49        & power & pri & 3.6 & 2.0  & 1.0  &  3   \\  
ESO H$\alpha$ 282 & power & sec & 6.6 & 1.5  & 1.5  &  1   \\  
B59-1             & phase & pri & 47  & 1.2  & 47   &  2   \\  
B59-2             & power & pri & 2.3 & 1.1  & 1.1  &  1   \\  
\enddata
\tablecomments{ 
Fits were made to the Fourier phase data for the two fully-resolved doubles, and to the power spectra for all other targets.
$\chi^2_{u}$, $\chi^2_{s}$, and $\chi^2_{h}$ are the $\chi^2$ per degree of freedom for the unresolved, double, and halo models, respectively.
The comment numbers are:
(1) Marginally-resolved, consistent with a tertiary or a halo.
(2) Fully-resolved double.
(3) Marginally-resolved, consistent with a halo but not a tertiary.
(4) No evidence of resolved structure
(5) Infrared Companion.  This system is also known as DoAr 24E.
}
\end{deluxetable}

The J,H,K,L' photometric measurements for the 14 holographic imaging targets
are given in Table \ref{PhotSummary}.  Measurements of the separation
vectors of their components, which are the averages of those measured at J,
H, and K with the corrected platescale, are included in the table.  The
brightness ratios for the visual binaries are plotted in Figure
\ref{BrightnessRatios}.

\begin{deluxetable}{lrrrrrrrrrrr}
\footnotesize
\tablecaption{ Infrared Photometry and Seeing-Limited Imaging
\label{PhotSummary}}
\tablewidth{0pt}
\tablehead{
\colhead{Object}
 & \colhead{J$_{pri}$} & \colhead{H$_{pri}$} & \colhead{K$_{pri}$} & \colhead{L'$_{pri}$}
 & \colhead{J$_{sec}$} & \colhead{H$_{sec}$} & \colhead{K$_{sec}$} & \colhead{L'$_{sec}$}
 & \colhead{Sep (\arcsec)} & \colhead{PA (deg)}
}

\startdata
    Sz 116				& 10.5  &   9.5  &  9.9  &        &  11.1  & 10.4  &   10.4  &	     &  1.65 & 26.5  \\
    As 205				&  8.3  &   6.9  &  6.2  &   5.4  &   8.8  &  7.5  &    7.0  &   5.9 &  1.34 & 211.3 \\
     WSB 4				& 11.6  &  11.0  & 10.9  &  10.1  &  11.9  & 11.2  &   10.6  &   9.1 &  2.84 & 128.5 \\
    WSB 18				& 11.1  &  10.1  & 10.0  &   8.9  &  11.8  & 10.8  &   10.4  &   9.0 &  1.08 & 80.4  \\
    WSB 19				& 10.8  &   9.7  &  9.6  &   8.4  &  11.5  & 10.5  &   10.4  &   9.3 &  1.53 & 260.7 \\
    WSB 26				& 10.9  &   9.9  &  9.6  &   8.5  &  11.4  & 10.2  &    9.7  &   8.1 &  1.15 & 23.8  \\
Elias 2-22				&  8.7  &   7.5  &  7.1  &   6.1  &  11.6  &  9.1  &    7.7  &   5.6 &  2.05 & 148.6 \\
    WSB 35				& 10.8  &   9.7  &  9.3  &   8.0  &  11.5  & 10.7  &   10.6  &   9.9 &  2.29 & 130.3 \\
Elias 2-26				& 10.5  &   8.8  &  8.4  &   7.2  &  11.9  & 10.2  &    9.6  &   8.3 &  1.44 & 67.0  \\
    WSB 71				& 10.1  &   8.8  &  8.0  &   6.2  &  11.0  & 10.1  &    9.9  &   9.2 &  3.56 & 35.0  \\
Elias 2-49				&  7.1  &   6.4  &  5.9  &   5.2  &   8.5  &  7.8  &    7.8  &   6.8 &  1.09 & 221.7 \\
ESO H$\alpha$ 282 &  9.8  &   8.6  &  7.8  &   6.8  &  11.1  & 10.1  &    9.5  &   8.6 &  1.03 & 251.8 \\
     B59-1				& 11.8  &   9.8  &  9.0  &   7.5  &  13.1  & 11.5  &   11.0  &  10.2 &  3.44 & 109.3 \\
     B59-2				& 10.0  &   9.3  &  9.3  &   8.9  &  11.3  & 10.4  &   10.1  &   8.9 &  3.11 & 207.0 \\
\enddata
\tablecomments{ 
  Near-infrared magnitudes and separation vectors for the holographic target
  binaries, based on corrected JHK measurements as described in the text. 
  The position angles are computed for vectors extending from the brighter
  to the fainter component in the J filter.  They are measured in degrees
  east of North.  The estimated photometric uncertainties are typically 0.2
  mag, and the scatter between the J, H, and K separation vector
  measurements for a given star is typically smaller than 0\arcsec.01 in
  separation and 0.2 deg in position angle.
}
\end{deluxetable}

Out of the 14 visual binary systems (nominally 28 stars) observed, two are
found to be well-resolved doubles, 5 to be marginally resolved along a
single axis (consistent both with doubles with separations less than the
diffraction limit, or with elongated halos), and two were
marginally-resolved along all axes.  The results for the individual resolved
and marginally-resolved sources are discussed below.


\subsection{ Fully-Resolved \\ Primary-Tertiary Pairs }

Two systems (WSB~18 and B59-1) have well-resolved tertiary companions
separated from their primaries by $\sim 0$\arcsec.1.  Models were fitted to
their Fourier phases in order to derive the brightness ratio $R$, separation
$\rho$, and position angle $\theta$ of the primary-tertiary pairs.  The
results are given in Table \ref{BinaryFits}.  

\begin{deluxetable}{lrrr}
\footnotesize
\tablecaption{ Double-Star Fits 
\label{BinaryFits}}
\tablewidth{0pt}
\tablehead{
\colhead{Object} & \colhead{$\theta$} & \colhead{$R$} & \colhead{Sep} \\
\colhead{} & \colhead{(deg)} & \colhead{}  & \colhead{(mas)}
} 
\startdata
Sz 116		      	 & $143.3  \pm .4$ &   $0.37   \pm .01 $  & $36.9  \pm  .4 $  \\
AS 205						 & $101    \pm 1 $ &   $1      \pm .003$  & $8.5   \pm  .4 $  \\
WSB 18		      	 & $339.55 \pm.05$ &   $0.505  \pm .001$  & $100.4 \pm  .1 $  \\
Elias 2-22      	 & $ 44    \pm 3 $ &   $0.98   \pm .001$  & $5.6   \pm  .2 $  \\
ESO H$\alpha$ 282  & $16.7   \pm .6$ &   $0.78   \pm .02 $  & $12.8  \pm  .1 $  \\
B59-1 		      	 & $218.9  \pm .1$ &   $0.49   \pm .002$  & $106.7 \pm  .1 $  \\
B59-2 		      	 & $113.5  \pm .5$ &   $0.91   \pm .05 $  & $12.2  \pm  .1 $  \\
\enddata
\tablecomments{ 
Double-star fitting parameters for the Fourier phase data on the two objects
with clearly-resolved tertiary companions, and fits to the Fourier power
spectra of the five objects which exhibit marginally-resolved structure
which consistent with the double-star model.  In the latter cases, the $R$
(brightness ratio) and $\rho$ (separation) values must be considered
representative of a class of possible models, and the position angle
($\theta$) values are ambiguous by 180 deg.  The errorbars quoted here are
derived from the standard deviation of the mean derived from the scatter
among fits to $\sim 7$ series of $\sim 140$ frames each.  They do not
include any potential contributions from uncertainties in the platescale or
orientation of the camera, or noise biases. They are very optimistic,
especially for the marginally-resolved objects.
}
\end{deluxetable}

The presence of a tertiary companion to the primary in WSB~18 is
noteworthy in light of the study by Brandner \& Zinnecker (1997), in
which the primary appeared somewhat more luminous than would have been
predicted for a single star with its M2 spectral type and the $5 \times
10^6$ yr age derived by comparing the secondary to the theoretical
pre-main sequence tracks of D'Antona \& Mazzitelli (1994).  It is natural
to suggest that the tertiary companion is responsible for the excess
luminosity, and its detection strengthens the conclusion reached by
Brandner \& Zinnecker (1997) that the components of the binaries in their
sample are coeval.

The projected separations of these primary-tertiary systems correspond to
$\sim 16$~AU at 160~pc.  Assuming that the true separations are not much
larger, that the mass of each close pair is $\sim0.6~{\rm M}_\odot$, and
that the orbits are circular, results in an estimate of $\sim 10^2$ yr for
their orbital periods.  The systems should rotate by $\sim 4$ deg yr$^{-1}$,
making their orbital motion readily detectable from one year to the next.

\subsection{ Marginally-Resolved Sources }

As discussed above, the nature of the 7 \\ marginally resolved targets is
somewhat ambiguous.  Five of them appear to be extended along only one axis,
making them consistent with both the double-star model (the double having a
separation below the diffraction limit) and the halo model (the minor axis
being much smaller than the major axis).  In these cases, the double-star
model is preferred for its astrophysical simplicity, although it may be
possible to produce an elongated halo of the observed size by a suitable
arrangement of circumstellar material such the inner regions of a nearly
edge-on disk.  The results of fitting the double-star model to these sources
are included in Table \ref{BinaryFits}.

\begin{deluxetable}{lccrrr}
\footnotesize
\tablecaption{ Gaussian Halo Fits to Multiaxial Power Spectra 
\label{GaussianFits}}
\tablewidth{0pt}
\tablehead{
\colhead{Object} & \colhead{comp} & \colhead{Fit to} & \colhead{$\theta$} & \colhead{$a$} & \colhead{$b$} \\
\colhead{} & \colhead{} & \colhead{} & \colhead{(deg)} & \colhead{(mas)} & \colhead{(mas)} 
}
\startdata
WSB 71						& pri & power &  $8   \pm 3$  & $64 \pm 1$ &  $50 \pm 1$  \\
Elias 2-49      	& pri & power &  $116 \pm 3$  & $54 \pm 1$ &  $45 \pm 1$  \\
\enddata
\tablecomments{ 
Gaussian-halo fitting parameters for the Fourier power data on the two objects which are
marginally resolved but not consistent with the double-star model.  The
parameters are the position angle of the major axis and the lengths of the
semimajor and semiminor axes.
}
\end{deluxetable}

The power spectrum of the Sz~116 secondary is consistent with a double whose
separation is just below the diffraction limit.  The best-fit model gives a
projected separation of 40 mas, corresponding to 6 AU at 160 pc, making this
the widest of the potential marginally-resolved doubles in the sample.   
The secondaries in AS~205, Elias~2-22, ESO~H$\alpha$~282, and the primary in
B59-2 also have asymmetrical power spectra consistent with the presence of
tertiary companions with separations ranging from only 6 to 13 mas, well
below the diffraction limit.  If the structures seen in these targets do
reflect tertiary companions, their periods would be only $\sim 1-20$ yr. 
For these marginally-resolved objects, followup holographic observations to
search for significant changes in the position angles of their extended
structures would provide a good test of the tertiary-star interpretation.


The primary components in Elias~2-49 and WSB~71 are marginally-resolved but
inconsistent with the double-star model because there is no position angle
along which the power spectra do not decrease with frequency.  In these
objects the $\chi^2$ is lower for the halo model than for the double-star
model.  The data are insufficient to determine the morphology of these
sources in any detail, and it may be much more complex than a simple
Gaussian halo.  The marginally-resolved primaries in these systems are much
redder than the unresolved secondaries (see Figure \ref{BrightnessRatios}),
which is not surprising if they are surrounded by dust envelopes, provided
that the overall flux is not dominated by scattering.  These sources merit
followup observations at shorter wavelengths, {\it e.g.,} visible-light
imaging with the {\it Hubble Space Telescope}, to search for extended, low
surface brightness scattered light to which the present holographic
observations are insensitive.



\subsection{ Wavelength Dependence of Visual Binary Separation Vectors }

If a star has a close companion with a significantly different infrared
color, or if it is surrounded by an asymmetrical halo of non-gray dust, then
its photocenter will be wavelength-dependent.  A comparison between the
separation vectors measured for a visual binary over a range of wavelengths
can potentially detect this effect.   The RMS scatter in the differences
between the IC-in and IC-out separation values for the binaries, excluding
Sz~116 (which was not measured at L') and Elias~2-49, is 0\arcsec.014.  This
gives an indication of the precision of the individual separation
measurements.  For Elias~2-49 the separation decreases with wavelength from
1\arcsec.09 at J to 0\arcsec.97 at L', a change which is much larger change
than the estimated precision and suggestively similar to the estimated
Gaussian size of the halo surrounding the primary star.  Deeper
high-resolution images of this system at visible or near-infrared
wavelengths would be useful to reveal the structure of the circumstellar
material.  Interestingly, WSB~71, the other system which appears to have a
halo around its primary, has a separation which is not dependent on
wavelength between the J and L' bands.

\subsection{ The Density of Tertiary Companions }

If one assumes that all five marginally-resolved sources which are
compatible with doubles do have physical tertiary companions at the
separations derived from the model fits, then fully half of the nominal
visual binaries in the sample are actually hierarchical triples.  This
perhaps surprising result is discussed below.

The angular surface density $\Sigma(\theta)$ of companions to young stars in
the Taurus, Ophiuchus, and Orion star-forming regions was computed by Simon
(1997) for companion separations $\theta$ as small as 0\arcsec.01.  In the
``binary regime" in which $\theta$ is small enough that $\Sigma$ is
dominated by bound objects, the surface density in all three clusters was
found to be $\Sigma(\theta) \propto \theta^{-b}$, where b is close to 2.0. 
A similar conclusion was reached for an X-ray selected sample of Taurus
stars by Kohler \& Leinert (1998; hereafter KL98) for separations between
0\arcsec.13 and 13\arcsec.  The robustness of this result is impressive in
light of range of physical conditions spanned by the three SFRs.

\bigskip
\begin{figure*}[htbp]
\epsfxsize=10cm
\epsfbox{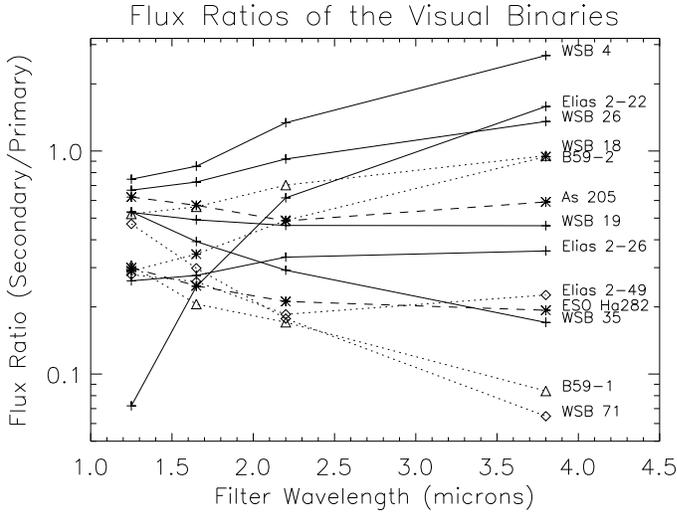}
\caption[ ]{ The brightness ratios for the visual binaries in the sample,
estimated by fitting the seeing-limited images with models consisting of a
pair of circular Gaussians, for the four filters.  Sz 116 is not included
on the plot because the L' measurement on it was not successful. }
\label{BrightnessRatios}
\end{figure*}

\bigskip
\begin{figure*}[htbp]
\epsfxsize=10cm
\epsfbox{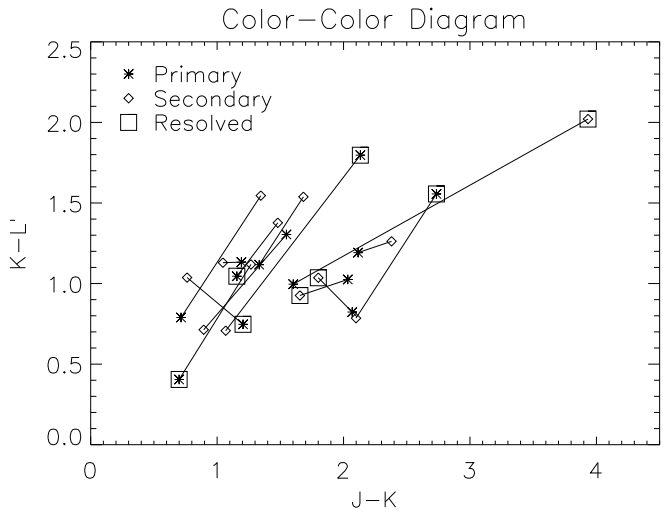}
\caption[ ]{ The near-infrared colors of the components of the visual
binary systems in the sample are plotted with symbols indicating their
rank ({\it i.e.,} whether they are primary or secondary stars).  Symbols
for the components of each system are joined by a line.  A box surrounding
a symbol indicates that it was marginally or fully resolved by the
holographic observations.  The data reveal no clear correlation between
brightness, color, or structure. }
\label{ColorColor}
\end{figure*}

Suppose that this result applies within the present sample of visual
binaries for tertiaries at separations small compared to the visual binary
separation.  This cannot be strictly true, both because the sample is {\it
statistically} biased by the presence of the visual companions and because
the companions' gravitational influence restricts the range of separations
for which a stable tertiary orbit may exist, but given the large ratio of
the observed outer to inner separations and the steepness of the density
function, it may not be terribly wrong.  The orbital stability restriction
has been estimated by Harrington (1972) and depends on the inclination of
the inner orbit relative to the outer orbit, the eccentricity of the outer
orbit, and the mass ratios of the components, but for an outer-orbit
eccentricity of 0.5 and components of similar mass it requires the major
axis of the outer orbit to be approximately a factor of 10 larger than that
of the inner orbit.  The average projected separation of the visual binaries
in the present sample is 2\arcsec, so according to the Harrington criterion
with the assumed parameters, and neglecting the effects of projection onto
the plane of the sky, the systems on average should permit tertiary
companions with separations no larger than about 200 mas.  Normalizing
$\Sigma$ to produce the number of companions observed by KL98 within the
range of radii to which their observations were sensitive, and integrating
over an annulus extending from the approximate detection limit of 5 mas to
the stability cutoff at 200 mas, leads to a prediction that 11 tertiaries
should be found among the 28 primary and secondary stars in the sample.  
Given the small sample size, the crudeness of the above approximations, and
the possibility that some additional tertiary companions were missed, this
is consistent with the 7 detected here.

\subsection{ Infrared Brightness Ratios }

The brightness ratios for the target visual binaries measured in all four
filters, plotted against the filter wavelengths in
Figure~\ref{BrightnessRatios}, show no clear tendency for the primary star
in a visual binary to be redder than the secondary.  Such a tendency would
be manifested as a generally negative slope in the brightness-ratio curves. 
This contrasts with the result for the sample of Taurus visual binary TTS
observed by Moneti \& Zinnecker (1991), in which the primary stars were
redder than the secondaries.  An examination of the color-color plot in
Figure~\ref{ColorColor} shows no clear correlations between the colors of
the visual binary components, their structure in the holographic
observations, or their relative brightness.

\section{ Conclusions }

The present study is among the first to probe a sample of pre-main sequence
visual binary systems with this combination of high resolution and dynamic
range in the near-infrared.  The results are not very consistent with the
naive expectations that motivated it.  More than half of the systems contain
resolved or marginally-resolved components.  Two tertiary companions were
clearly resolved at $\sim 100$ mas separations from their primaries, and
tertiaries at separations below the diffraction limit are likely to account
for the structure seen in all but two of the marginally-resolved stars,
including the infrared companion (IRC) in Elias~2-22.  Followup observations
to search for orbital motion would be useful.  The presence of such a large
number of tertiary companions would make it necessary to search binary
systems carefully before attempting to use them to test pre-main sequence
tracks, since the light of an undetected tertiary will make a star appear
younger and more luminous that it is.  Two stars showed extended structure
which appears to be due to the presence of dust halos rather than tertiary
companions, and one of these (Elias~2-49) has a separation which decreases
with wavelength.  Strikingly absent is a clear detection of an edge-on
circumstellar disk in any of the binaries in the sample.


\acknowledgments

It is a pleasure to thank M. Creech-Eakman for her assistance with the
observations, G. Blake, P. Goldreich, and J. Kwan for useful discussions, E.
Gaidos for useful comments on a version of the manuscript, and to an
anonymous referee for an exceptionally useful report.  Data presented herein
were obtained at the W.M. Keck Observatory, which is operated as a
scientific partnership among the California Institute of Technology, the
University of California and the National Aeronautics and Space
Administration.  The Observatory was made possible by the generous financial
support of the W.M. Keck Foundation.  The author wishes to extend special
thanks to those of Hawaiian ancestry on whose sacred mountain he was
privileged to be a guest.  Without their generous hospitality, and the hard
work and dedication of the Keck Observatory staff, the observations
presented herein would not have been possible.  This research was supported
by the National Aeronautics and Space Administration.



\clearpage

\clearpage

\end{document}